\documentclass[prl,twocolumn,showpacs,floatfix,amsfonts]{revtex4}
\usepackage{graphicx,graphics,color,epsfig}% Include figure files
\usepackage{bm}
\usepackage{amsmath}
\usepackage{amssymb}

\newcommand{\beqa}{\begin{eqnarray}}
\newcommand{\eeqa}{\end{eqnarray}}

\begin{document}
\preprint{}
%\title{Q Balls and Bose-Einstein Condensates}
\title{Non-Relativistic Bose-Einstein Condensates, Kaon droplets,
and Q-Balls}

\author{Zohar Nussinov}
\email{zohar@viking.lanl.gov}
\affiliation{Theoretical Division, Los Alamos National Laboratory,
Los Alamos, New Mexico 87545 }

\author{Shmuel Nussinov$^{(a)}$}
\email{nussinov@post.tau.ac.il}
\affiliation{Department of Physics and Astronomy,
University of South Carolina, Columbia, USA\\
$^{(a)}$on sabbatical leave from School of Physics and
Astronomy, Tel Aviv University, Ramat-Aviv, Tel Aviv 69978, Israel}

\begin{abstract}
       We note the similarity between BEC (Bose-Einstein
Condensates) formed of atoms between which we have long-range
attraction (and shorter-range repulsions) and the field
theoretic "Q balls". This allows us in particular to address the
stability of various putative particle physics Q balls made of
non-relativistic bosons ($K^0$'s, $B^0$'s, and $D^0$'s) using
variational methods of many-body NRS (Non-Relativistic
Schr\"odinger) equation.

\end{abstract}

\maketitle

\subsection{Introduction}

  Phase transitions occur when a change of coupling or
 temperature makes the (free) energy of a new phase lower than
 that of the preexisting phase. In the field theoretic
 formulation, this is manifest when the minimum of the effective
 potential $U(\phi)$, with the field $\phi$ representing some order
 parameter, is shifted from $\phi=0$ to nonzero $\phi$ value or to a
 degenerate manifold of such $\phi$ values. In a simple mechanical
 analog the system represented by the single coordinate $\phi$,
 ``rolls over" to the new stable minimum. Some time ago, Sidney
 Coleman introduced\cite{Coleman} ``Q balls": classical field
 configurations stabilized by the global conserved charge (s) (Q)
 they carry. These correspond to configurations of, say, a charged
 field which are constant over a large region of space but vary
 in time: $[\phi ~ \exp(-iw_0t)]$ and $[(\phi)^+ ~ \exp(iw_0t)]$---
 corresponding to a constant charge density $j^0=\rho =
 (\frac{d}{dt}\phi^+ \cdot \phi - \phi^+ \frac{d}{dt}\phi)/2i$. The simple
 mechanical analog here is a system rotating with uniform
 angular velocity in the $\phi_1 - \phi_2$ plane (with $\phi =
 \phi_1 + i \; \phi_2)$. The ``centrifugal" force generated can
 then, under certain conditions specified below, make the
representing ``particle"  come to equilibrium at a $\phi_0$ which is
 no longer a minimum of $U(\phi)$. Coleman's suggestion 
of using the conserved baryon number as the global charge of
 the Q balls has been followed up in super-symmetric
models\cite{Kusenko}. Motivated by advances in BECs 
(Bose-Einstein condensates), we
 investigate in this paper their relation to Q balls and use
 variational NRS (non-relativistic Schr\"odinger) equation methods
 to prove the stability against strong interaction decays of
 Strangeness, Charm, and Beauty balls.

 Field theoretic Q balls are more general than the
 non-relativistic limit on which we focus here. Thus\cite{Kusenko}
 the above-mentioned baryonic Q balls, made of squark condensates,
 are stable against ``weak interaction like" decays of the heavy
 individual squarks by having the Q balls very tightly bound with
 masses proportional to a fractional power of the
 total baryon number $N^{(1-\epsilon)}$ rather than $N^1$ as
 expected for non-relativistic weakly bound matter.

  Still, the equivalence---in some limits---of the
  field-theoretic and many-body descriptions of the same
  ``condensations" is interesting and helpful.

  There are {\it two} distinct, though interrelated, types of
 condensation. BECs obtain when a large number, $N$,
 of (bosonic) atoms are trapped within a  radius $R $ and then
 cooled down to nano-Kelvin temperatures. The BE condensation is
 in {\it momentum} space: a finite fraction  of the atoms are in
 the lowest mode of the trap corresponding to $p=0$. This
 manifest when the trap is suddenly removed by having these $p=v=0$
 atoms hardly move.\cite{Saito} Boson-boson interactions modify
 the BEC but are {\it not} responsible for BEC phenomenon in the
 first place. BEC is best understood when the atoms are
 non-interacting\cite{Pitaievsky}. On the other hand the {\it
 coordinate} space ``condensation" of bosons into Coleman's ``Q
 balls" is due to attraction between the bosons. Thus in the
 field theoretical formulation Coleman has proved that stable Q
 balls exist if and only if the``potential" $U(\phi)$ in the effective
 low-energy Lagrangian for the system satisfies:
\begin{equation}
{\rm (i)} \;\;\;\;\;\;\;\;\;\;\;\;\;\;\;
\frac{U(\phi)}{\phi^2} \; {\rm has \; a \; minimum
\; lower \; than} \; \frac{\mu^2}{2}
\label{Criterion(i)}      %{Eq 1= Criterion (i)}
\end{equation}
with $\mu$ the mass in the ``free" part of $U(\phi):\; \frac{\mu^2}{2}
\phi^2$.

 The size $R$ of the spherical Q ball and its density 
$n=(3Q)/(4 \pi R^{3})$ are fixed by the overall 
Q and $\phi_0$---the field
 for which the above minimum is achieved.

  Condition (i) implies over all attractive interactions between
  the $\phi$ bosons, say, the kaons in strangeness balls, at an
  appropriate density. In particular, (i) holds if the
  coefficient of the lowest $(\phi)^4$ term in $U(\phi)$ is
  negative:
\begin{equation}
{\rm (ii)} \;\;\;\;\;\;\;\;\;\;\;\;\;\;\;
U(\phi) = \frac{\mu^2}{2} \phi^2 - \lambda
\phi^4 + \; {\rm higher ~ order \; terms}
\label{Condition(ii)}    %{ Eq. 2 = Condition (ii)
\end{equation}
corresponding to attractive S-wave scattering length
($\lambda > 0$ is implicit). Note that  $U(\phi)$ is an {\it effective}
Lagrangian, which is {\it not} used in loops inside Feynman
diagrams. Therefore $U(\phi)$ can have (and indeed has) higher-order
{\it positive} non-renormalizable $(\phi^+ \phi)^n$
terms-ensuring a finite $\phi_0$ and a spectrum which is bounded
from below.

 While (ii) $\rightarrow$ (i), namely, an attractive S-wave
scattering length (negative $\phi^{4}$ coefficient), implies stable Q balls,
(ii)  is {\it not}  required. Thus, a nontrivial minimum, $\phi_0$,
of $U(\phi)/(\phi)^2$ can obtain with a negative $(\phi^+
\phi)^3$ term overcoming at some $\phi$ the positive $ \frac{\mu^2}{2}
(\phi)^2 + \lambda (\phi)^4$. Still this can be problematic: the
minimum of $U(\phi)/(\phi)^2$  may now be at a large $\phi_0$,
say, $\phi_0 >> \mu$ where the NR many-body approach that we
want to compare with next, fails.

 Note that Condition (ii) alone, without knowledge of the
 higher-order terms,  does {\it not} fix the size ($R$) or the density
 ($n$) of the Q balls: with only the $[-\lambda \phi^4]$ term
 present both $ \phi_0$ and n are infinite!

%II.
\subsection{Forming a Spatial Droplet of Non-relativistic
Bosons: The Many-Body Approach}

  Let $N$ non-relativistic identical bosons of mass $m$ interact
via potentials $V(|r_i-r_j|)$. The basic question we address 
is: ``For which potentials the NR bosons coalesce into ``Q balls" with
nonzero density when $N \rightarrow \infty$ ?"

 For finite range potentials the N-body Schr\"odinger equation
 $H|\Psi \rangle = E|\Psi \rangle$ with
\begin{equation}
  H=\sum_{(i)} \frac{p_i^2}{2m} + \sum_{(i>j)} [V(|r_i-r_j|)]
\label{finiterange}   %    {Eq .3}
\end{equation}
is trivially solved by $\Psi$ which is a  product of single particle
wave functions, each of which  is (approaching) a constant.
The spread-out particles have vanishing kinetic energies and
vanishing mutual interactions and thus $E=0$. For many potentials a
lower (negative) energy state exists with the $N$ bosons in a
sphere of radius $R$. Using the basic variational principle we
derive next several {\it sufficient} conditions for that to
happen.  In Secs. III,IV  below we argue that some of these sufficient
conditions are met in the case of many $K^0$'s, $D^0$'s and $B^0$'s
which therefore will make droplets or Strangeness, Charm and
Beauty balls which are strong interaction stable.

Let us first use a simple trial wave function with all bosons in the same
 state---the ground state of a large radius $R$ spherical cavity:
\begin{equation}
    \Psi_t = \prod_i(\psi^0(r_i)).
\label{trialwf}   %  {Eq. 4}
\end{equation}
The expectation values of the kinetic and potential energies appearing in
 $ \langle \Psi_t|H|\Psi_t \rangle   =  
\langle K \rangle  +  \langle V \rangle $
 are:
\begin{equation}
\langle K \rangle \simeq \, N  \hbar^2/(2m \, R^2)
\;\;\;\;\;\;\;\;\;\;\;\;\;\;\; (a)
\label{expvalues5a}
\end{equation}
and, if $R >>  r_0$ = the range of the potential, 
$ \langle V \rangle  \simeq  {\rm [Nnv]} \sim [N^2/(4 \pi R^3/3)] v$, with 
\setcounter{equation}{4}
\begin{equation}
v  = 4 \pi \int \; V(r) ~ r^2 dr
\;\;\;\;\;\;\;\;\;\;\;\;\;\;\;\; (b)
\label{expvalues5b} % {Eq. 5a-b}
\end{equation}
and n=N/(volume) the particle number density.
A possible relation to the previous section stems from the fact
that, up to kinematic factors, $v$ is the Born approximation for
the S-wave boson-boson scattering length. For dilute systems with
inter-particle separation which far exceed the range of the
potential:
\begin{equation}
         d = n^{-(1/3)} >> r_0 \;,
\label{dilsys}  %    {Eq.6}
\end{equation}
only the integrated potential $v$ effects the threshold scattering.
The threshold scattering amplitude is therefore reproduced also
by a ``pseudo-potential" of a local delta function form: $v
\delta^3(r)$. Such a potential roughly corresponds to the
negative local $-\lambda (\phi^4)$ term---which in the field
theoretic formulation suffices to ensure stable Q balls. To
complete the analogy with the NRS case we show that if $v < 0$,
then also the expectation value of the energy $ \langle H \rangle < 0$. This
readily follows from the different scalings of  the
(expectation values of) the positive kinetic and negative
potential energy $N$, the number of bosons in the system: $ \langle K 
\rangle \sim
N^{(1/3)} << | \langle V \rangle| \sim N$ for $N \rightarrow\infty$. By the
variational principle the energy of the true N-body ground state
is lower than $ \langle H \rangle $ and also negative and a many-body bound
droplet or Q ball stable state indeed exists if:
\begin{equation}
  (ii'):    v =  4 \pi \int V(r)~ r^2 dr < 0  \rightarrow {\rm a \;
``droplet" \; state \;  exists}.
\label{Criterion(ii)'}  %{Eq.7= Criterion (ii)'}
\end{equation}
 Like (ii),  (ii$'$) does {\it not} fix the actual size/density of the
``droplet".
 Still the two conditions are {\it not} equivalent.

The quantity $v$, depends only on the potential $V(r)$ and not
the mass $m$. It is (proportional to) the actual S-wave
scattering length (or to the coefficient of the $\phi^4$ term in
$U(\phi)$ {\it only} in the Born (or dilute system) approximation.

 Scattering  theory\cite{Regge}
 implies that the S-wave scattering length is attractive if and only if the
 NRS (non-relativistic Schr\"odinger) two-body system has bound
 states. Thus the NRS equivalent of the field theoretic condition (ii) is
 having a two-body bound state of the NRS equation. Indeed as we
 directly show below having two-boson bound states guarantees $N$
 bosons bound state.
 Note, however, the independence of the two NRS criteria: The
criterion (ii$'$) ensuring N $\rightarrow \infty$ NRS ``droplets"
does {\it not} ensure an S-wave two-body bound state. In three
dimensions the latter requires not only an ``attractive"
potential, but also sufficiently strong attraction.

Various criteria for $V(r)$ to have bound states in a NR two-body
system with reduced mass $m$ exist.\cite{Bargman} Yet there is no
general {\it if and only if} criterion, short of solving the Schr\"{o}dinger
equation. Finding if a bound state of $N \rightarrow \infty$
bosons exists need not be easier. Indeed the field theoretic
Criterion (i) requires the full effective potential $U(\phi)$. The
coefficients in the power series for the latter are the threshold
scattering amplitudes for any number of particles and cannot be
found short of solving exactly the field theory.

%III
\subsection{``Strangeness-Beauty Balls" and the
Non-Relativistic Schr\"{o}dinger Equation}

  For many atomic and other systems, the many-body NRS treatment
 preceded field theoretic approaches and Laughlin's celebrated
 variational wave function for the fractional quantum Hall effect
 is a prime example of this. In recent years much effort has
 been devoted to applying field theory in general, and effective
 field theories and effective Lagrangians in particular, to such
 problems and this has been also the case for the
 Q balls.\cite{Enquist} %CERN}

  Chiral Lagrangians were used to check if Coleman's Criterion
 (i) holds for the $K^0$ system. These Lagrangians
 involve higher derivative  terms and are fixed by an overall fit to data
 yet they do not offer much intuitive understanding. Here we
 follow the {\it reverse} program of ``demystifying" Q balls and
 trying to explain them---at least in the NRS regime---as simple
 spatial condensations of many bosons with appropriate
 potentials. Strangeness-balls, and even more so,
 Charm/Beauty-balls with densities $n<m_K^{-3}$ are NR. In  all cases
 we expect to have essentially the same  boson-boson potentials as the
 later are controlled by the common light $d$ quark. Hence, using
 $K^0-K^0$ potentials, $V_K(r)$, etc., to find if ``$K^0$, etc.,
 Droplets" form---which we do next---is justified.

  A first key observation is that the $K-K$ potential, as that
  between {\it any} two identical, neutral, pseudo-scalars, is
  {\it  attractive} at ``large" $\sim$ Fermi distance.
To show this we write $V(r)$ as a superposition (integral) of
 Yukawa potentials due to all exchanges\cite{Au}:
\begin{equation}
   V(r) \simeq - \int d \mu ~\sigma(\mu^2)~ \frac{e^{-\mu  r}}{r}
\label{VrIntegral}  %   { Eq. 8}
\end{equation}

  Parity conservation forbids a $KK\pi$ vertex and the lightest
exchanged system controlling $V(r)$ at $r \rightarrow \infty$
is that of two pions. Further, the $\ell=0$ component dominates at
the $\pi\pi$ threshold:
\begin{equation}
 \sigma (\mu^2) \; {\rm near} \; \mu =2 m(\pi)
\sim |f(l=0) (K+\bar{K} \rightarrow \pi \pi)|^2
\label{l=0component}  %  {Eq. 9}
\end{equation}
With $f(\ell=0)$, the $\pi \pi \leftrightarrow K\bar{K}$ S-wave
amplitude. This expression is clearly positive and recalling the
minus sign in the definition of $V(r)$ we find that the longest range
two-pion exchange potential is indeed attractive. This can be
also directly shown to be the case for the exchange of a $0^{++}$
state which can be an S-wave resonance in the $\pi \pi$ S-wave
system. The above reasoning is similar to that used in \cite{Au}
to derive the well-known attractive Casimir Polder (i.e retarded
van der Waals)  and the regular van der Waals two photon-exchange
potential between two identical neutral atoms.

  At short distances of the order of the size of the $s\bar{d}$
composite state, which {\it is}  the kaon in quark
models/QCD, the $KK$ potential becomes repulsive. As in atomic
physics this is due to the Pauli principle---operating here
between the identical $\bar{d}$ or $s$ quarks in the two $K^0$'s.
The repulsion can also be viewed as being due to the exchange of
the $\omega \; (\rho)$ vector mesons: The two $K^0$'s have the
same hypercharge (isospin) to which $\omega \; (\rho)$
couple\cite{Sakurai}; no light $0^{++}$, scalar ``$\sigma$" meson
has been established.  Yet, the exchange of such an entity with
appropriate mass and coupling to nucleons $g(\sigma;NN)$ (which
may represent the box diagram with two pions exchanged and
intermediate $N$ and $\Delta(1230)$ states) could, along with
$\omega$ exchange, dominate nuclear binding\cite{Walecka}.

 Parameterizing,
\begin{equation}
  V_K(r) = -g(\sigma)^2 \frac{e^{-(m_\sigma r)}}{r}
  + g(V)^2 \; \frac{e^{- m_V r}}{r}
\label{parametrizing}   % {Eq.10}
\end{equation}
  with $V =\omega$ or $\rho$ at a common mass $m(V)$ and $g(V)^2$ the
sum of
  the (squared) $\rho$ and $\omega$ couplings to kaons we find:
\begin{equation}
  v_K = - \frac{g(\sigma)^2}{m(\sigma)^2} 
+ \frac{(g(\rho)^2 + g(\omega)^2)}
{m(V)^2}> 0
\label{rhoomegacouplings}  % {Eq .11}
\end{equation}
To evaluate (\ref{rhoomegacouplings}) we take
$g(\rho,KK)=g(\omega,KK) \sim (1/3) g(\omega,NN)$
and $g(\sigma,KK) \sim (1/3)g(\sigma,NN)$,
 as suggested by counting the numbers of non-strange quarks.
 Using the values of $g(\omega,NN)^2/m_V^2$ and
$g(\sigma,NN)^2/m(\sigma)^2$
suggested by fitting nuclear matter in Eq. (14.27) in \cite{Walecka} 
we find that $v_K \sim (-2.35+3.45)/m_N^2 > 0$, so that Criterion
(ii$'$)
is {\it not} satisfied.  The box diagrams with $K^*$ intermediates for
$KK$
suggests a $g_{\sigma,KK}^2$ which is somewhat bigger than the previous
$[(1/9) g(\sigma,NN)^{2}]$ estimate. Also cutting off $V_K(r)$ of
Eq.(\ref{parametrizing})
in evaluating $v_K$ at, say, 0.3  Fermi---suggested by the fact that
the $K^0$
is composite at such a scale and can no longer be treated as a point
source of
the $\sigma, \omega$  fields---further reduces the repulsive relative
to the
attractive contribution and a negative $v_K$ is not excluded.

Still  $v_K > 0$ is likely and we face the question: ``Does a positive
$v_K$
exclude stable $K^0$ droplets?" This is the case if we insist on N-body
wave
functions which are products of $N$ identical {\it one}-particle wave
functions $\psi(r_i)$.  However, including the (Jastrow) product of $N
(N-1)/2$
{\it two}-particle functions
\begin{equation}
 \Psi_{(trial)} = \prod_i(\psi (r_i)) \prod_{(i>j)}  (f(|r_i-r_j|))
\label{Psi(trial)}  % {Eq. 12}
\end{equation}
we can have an N-body bound state even if $v>0$.

 To illustrate this we fix the potential $V_K(r)$ (and $v_K$)  and
increase the mass. (Taking $m= m_B \sim 11 m_K$ corresponds, in
the approximation of universal potentials between pairs of heavy
mesons---and treating $s$ as a heavy quark, to discussing ``Beauty"
(rather than Strangeness) balls.  While $v_K \sim v_B \sim v$
the system can now  have even two-body bound states.

 In the $m \rightarrow \infty$ limit these are localized around
the minimum of $V(r)$ at $r=r_0$, forming a ``Vibrational Band"
with  spacings $\sim([V''(r)|r=r_0]/m)^{(1/2)}$. Once the
potential $V(r)$ has two-body bound states, N-body bound states
are guaranteed. This is verified by using in Eq. (\ref{Psi(trial)})
$f(r)= \psi_0 (r)$ with $r=|r_i-r_j|$ satisfying:
\begin{equation}
 (- \frac{\hbar^2}{m} \frac{d^2}{dr^2} + V(r)) \psi_0(r) = E_0 \psi_0(r)
\label{Nbodybs}  %   { Eq 13 }
\end{equation}
with $E_0 < 0$ the negative energy of the two-body bound state,
and operating on the above trial function with the N-body
Hamiltonian of Eq. (\ref{finiterange}) above. (Here,
the reduced pair mass $m^{*}= m/2$ replaces the single particle mass 
in the Schr\"odinger equation.) 
Even when the
potential is too weak relative to the kinetic term to have a
two-body bound state, using Eq. (\ref{Psi(trial)}) with $f(r)$ of
the above general form, namely, peaking at the minimum of the
potential and being small at small $r$'s where $V(r)$ is
repulsive, lowers $ \langle H \rangle$ relative to its value 
for the product of
one-particle functions. As we
show in some detail in the next
section, this can yield the desired N-body bound state even when
Condition (ii$'$) fails, and also there are  no two-body bound
states. In this case we have a fully symmetric N-boson bound
state with the $ f(|r_i-r_j|)$ factors  peaking at $r=r_0$, and
determining the density of the N-boson droplet to be:
\begin{equation}
   n \sim ([(4 \pi)/3] r_0^3)^{(-1)}
\;\;\;\;\;\;\;\;\;\;\;\;\;\;\; (a)
\label{Nboxondroplet}  %      { Eq 14.a}
\end{equation}
or, equivalently, the radius of the droplet
\setcounter{equation}{13}
\begin{equation}
            R  \sim N^{(1/3)} r_0
\;\;\;\;\;\;\;\;\;\;\;\;\;\;\;\;\;\;\;\; (b)
\label{dropletradius}   %       {Eq 14.b}
\end{equation}

This is reminiscent of  $U(\phi)/(\phi)^2$ and its nontrivial
 minimum at $\phi_0$---fixing the radius and density of the Q
 balls in the field theoretic formulation. There the nontrivial
 $\phi_0$  obtains via the interplay between a negative $\phi^4$
 term and positive higher-order terms. In the present NRS case
 the minimum at $r_0$ reflects attraction (repulsion) at long
 (short) ranges. The higher $(\phi)^n$ terms are prominent at
 large densities---just like the strong short-range repulsions in
 NRS.

 Still, a $U(\phi) \leftrightarrow V(1/r)$ analogy is rather
limited: The {\it effective} potential $U$ derives from the
fundamental Lagrangian  of the field theory, say, QCD for the
above cases, whereas the potential $V(r)$ is the primary entity
in the NRS approach. A closer analog of $U(\phi)$ is the derived
quantity $[E/N](n)$---the energy per particle for a given
density\cite{WallacePC} in NRS. $E = E[N;R]$ is the ground state
energy of the Hamiltonian in Eq.(\ref{finiterange}) of $N =n [(4
\pi)/3] \cdot R^3$  bosons uniformly distributed (after averaging over
correlations) in a sphere of radius $R$.  When $N$ and $R$ tend
to $\infty$ keeping $n$ fixed, a stable droplet of density $n_0$
obtains that if and only if the minimum of [E/N](n) is  at
$0 < n_0 < \infty$ and  is  negative \cite{HolygrailFN}.

 % IV.
\subsection{Binding and BECs in the Presence of Strong, Short-Range
Repulsions}

   Short-range repulsions do not hinder BEC for dilute atomic systems
in traps.

Let N-bosons be in the trap and add one more. To see the issue
most clearly, assume first that the N-bosons are ``frozen" at
specific locations $r^0_i$ inside the trap. The added $ k<<N $
bosons will be in the ground state of the total potential:
\begin{equation}
 V= V_{(trap)}(r) + \sum_{(i=1,..N)} V(|r^0_i-r|)
\label{totalpotentialGS}  %{ Eq 16 }
\end{equation}
---provided that the potential \ref{totalpotentialGS} can bind a
particle of mass $m$
(which is clearly the case for an attractive $V_{(trap)}$ and
sufficiently dilute atoms).

 Conceivably such a setup, of interest in its own right, can be
experimentally realized.
Let the trapped $N$ atoms form a 3-D lattice generated by a
standing wave pattern of three lasers. Let the   added atoms
be of a {\it different} species which interact with the first $N$
atoms via the potential $V$ in Eq. (\ref{totalpotentialGS})
above. In particular, we need to choose a species which is
almost unaffected by the laser fields.

  ``Freezing"  $N$ out of $N+1$ is artificial and  adding one
 extra boson causes each of the previous $N$-bosons to adjust by
 order $1/N$, modifying the binding energy by O(1). A key point
 is that the  adjustments are likely to {\it lower} the energy
 and neglecting those is appropriate if we only want to verify
 that the extra particle binds.

 As the system becomes denser, stronger two-boson correlations
 and higher momentum components build up. The completely
 symmetric N-body state, while no longer factoring into
 independent N single-particle wave functions, still exhibits
 coherence features unique to  BECs.

 Our main interest however is not this, but rather the ``droplet
 formation" problem posed in the previous section. To address this
 problem  we use Eq. (\ref{totalpotentialGS})  without
 $V_{(trap)}$ and two-body potentials which are attractive
 (repulsive) at long (short) distances (Fig 1.)

\begin{figure}
\centerline{\psfig{figure=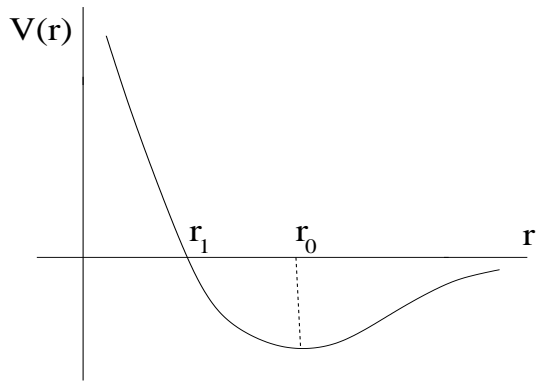,height=5cm,width=7cm,angle=0}}
\caption{The KK, DD  or BB NRS potential of interest.
It is repulsive at short distances $r < r_1$ and attractive for
$r >r_1$ having an asymptotic $\frac{e^{-2m_\pi r}}{r^3}$ behavior due to
two-pion exchange (qualitatively similar to a Lennard-Jones potential).  
The minimum of the potential is indicated by the dotted
line is at $r_{0}$.} \label{FIG:curve}
\end{figure}

A likely configuration the $N$ particles  $r^0_i$ is at
 the vertices of a simple cubic lattice of lattice constant
$d=n^{(-1/3)}$.
A simplified version of the problem which serves as a criterion
for formation
 of a droplet of density $n=d^{-3}$ is:
%\begin{itemize}
%\item[(iii)] 

(iii)``Does a particle  of mass $m$ bind to a  cubic lattice
with  lattice
constant $d$ and common potentials $ V(|r-r_i|)$ centered at all
lattice points
$r_i$?"
~~~~~~~~~~~~~~~~~~~~~~~~~~~~~~~~~~~~~~~~~~~~~~~~~~~~~~~~~~~~~~~~~~~~~~~~~~~~~~~~~%(17)
%\end{itemize}
%\label{Criterion(iii)}  %(Eq 17 )

---a problem which may also be  useful in discussing trapping of light
in some
``dielectric lattices" via the Helmholtz equation.

 It is difficult to address it in the most general case, yet the
following suggests that
 bindings are likely even when a single potential $V(r)$ fails to bind
and/or to satisfy
 $v<0$.
 Let  $r_0, r_1$ with $ r_0 > r_1$ be the points indicated in Fig. 1
where the common
 radial potential $V(r)$ has its minimum and where it changes sign,
being repulsive
 (attractive) for
$r< r_1 \; (r > r_1)$.  Consider a unit lattice cell of side of length
$d$ with eight potential centers at its corners. We focus first on the
``dilute" case with
$d>>2 r_0 >2 r_1$.
 Apart from the eight spherical octants of radius $r_1$ at the corners,
$V$ of
Eq. (\ref{totalpotentialGS})  above is attractive at all the remaining
part of the unit cell.
 The longer-range tails of the other
 potential centers make for an attractive, negative contribution in the
form of a
 ``Madelung sum" at any, (say, corner) point. This increases the volume
of the
connected region within the unit cell where $V$ (sum) $<$ 0 beyond the
minimal value:
\begin{equation}
  d^3 - \frac{4 \pi}{3} r_1^3  \;\;\;\;
[\sim .93 \; d^3 \;  if  \; d>4 r_1\;{\rm !}] \; .
\label{minvalue}  %{Eq. 18}
\end{equation}
 Thus we find that the particle is inside a ( very loose!) cage of size
$\sim d^3$
 ``cornered in" by the repulsive potentials centered at the eight
corners of the
 unit cell of the cubic lattice---as in a three-dimensional analog of a
carton egg holder.
 Actually, in the dilute case considered here, the particle can ``roam"
over all the
 lattice, going into neighboring cells  via the large circular openings
 of radius $d/2-r_1$ between neighboring cells, lowering the energy.
 To show this more clearly we simplify the problem by:
 (a) replacing the repulsive potentials within spheres of radius $r_1$
centered at the
 vertices of the lattice by infinite, positive,
``square-well" potentials within the circumscribing cubes of size $2
r_1$
centered---just like the above spheres---at the corners of the lattice
cell, and
 (b) replacing the attractive potential within the remaining region by
their volume
 average,
\begin{equation}
   -u_0=v^-/d^3
\label{replacingpotentials}     %     {Eq 19}
\end{equation}
 with $v^-$, the integral over the attractive part of the potential.
(See Fig. 2.)

\begin{figure}
\centerline{\psfig{figure=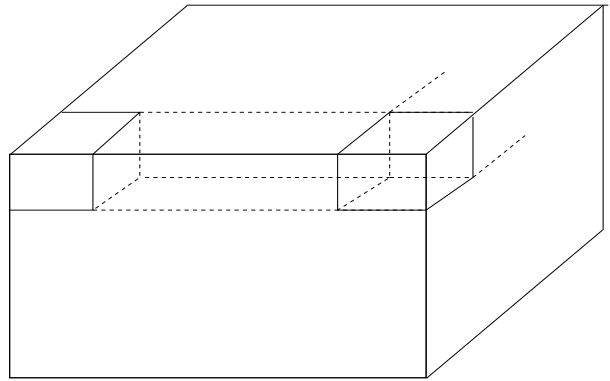,height=5cm,width=8cm,angle=0}}
\caption{ A schematic picture of the attractive and repulsive regions
(after
step b has been implemented) inside the unit cell---the overall cube of
side $d$.  The repulsive regions (with $V = +\infty$) are the eight
smaller cubes of
side $r_1$ at the eight corners of the unit cell---the big overall
cube. Three
of these are shown as the full line small cubes in the figure.
 The potential is attractive (negative) outside these eight cubes with
an
averaged constant value.
Using additional dashed lines we also illustrate two out of the twelve
 rectangular parallelepiped (RP) along the edges of the cube, between
 a pair of adjacent repulsive boxes. These can be viewed as
 ``wave guides" for the particle in the potential where we have
vanishing
boundary conditions on the small ($r_1^2$ in area) sides of the RP
 and the particle is propagating in directions perpendicular to the
long
($d-2 \; r_1$ in length) sides of the RP
 eg., in the top-front RP the wave guides in the up-down and
 forward-backward directions, and in the right-most RP illustrated in
the up-down
and the left-right directions.} \label{Fig1cube}
\end{figure}

 Step (a) increases the volume of the region where the potential is
repulsive by
 $6/\pi$ and the value of the potential which vanishes at its boundary
to $\infty$. This
 clearly increases the energy of the particle in the lattice.
 It is less obvious but still true that performing step (b), namely,
replacing the
 attractive part of the potential by its average value, also increases
the energy of the
 particle in the lattice. To see this, note that the true ground state
wave function for
 the original potential tends to concentrate away from the repulsive
regions and thus can
 sense better the attractive potential prevailing around the center of
the  above
 ``cage", i.e., unit lattice cell, and this attractive potential is
{\it stronger}
 before the averaging is performed.
 Hence, if after performing steps (a) and (b) above the particle still
binds to the
 lattice, namely,
 $(E_{(ground)} <0)$,
 then even deeper bound states are likely for the original problem.

 To test for a bound state we estimate the expectation value of the
 Hamiltonian $\langle H \rangle  = \langle V \rangle 
+\langle T \rangle$  in  the state $|\psi_{(ground)} \rangle$.
  Since the (normalized) wave function is nonzero only
outside the repulsive regions, the expectation value of the
constant attractive potential is:
\begin{equation}
   \langle V \rangle = -u_0 = v^-/d^3.
\label{expvalue}  %    { Eq .20}
\end{equation}

 The kinetic energy stems from the boundary conditions: $\psi=0$ at the
boundaries of the
 repulsive cubes. The twelve regions, of volume $ r_1^2 (d-2 \; r_1)$
each, inside the $d^3$
unit cell between pairs of repulsive corner cubes can be viewed as
short section parts of
``wave guides". The kinetic energy can be roughly approximated as that
of the lowest
 mode of the ``wave guides":
$ T = \hbar^2/[2 m (d-2 r_1)^2]$
 and its expectation $ \langle T \rangle$  
is weighted by the fraction of the unit
cell,
$ f = 12 r_1^2 \; (d-2 r_1)/d^3$, occupied by the ``wave guide"
sections, namely
\begin{equation}
 \langle T \rangle = f T = \frac{\hbar^2}{m}\, \frac{6 r_1^2}{d^3(d-2 r_1)}.
\label{waveguide} %   {Eq. 21}
\end{equation}
 The condition for binding, $\langle T \rangle +  \langle V \rangle < 0$ 
then becomes
\begin{equation}
 |v^-| >  \frac{\hbar^2}{m} \, \frac{6 r_1^2}{d-2 r_1}.
\label{bindingcond}   %               {Eq.22}
\end{equation}
 Since $v^-$ is independent of $d $, the last condition can clearly be
 satisfied for $d>>r_1$.

 As we gradually decrease $d$ (relative to the distance scales $r_0$
 and $r_1$ of the potential $V(r)$), the average value of the
 attractive part, namely, $|u_0|$ of Eq. (\ref{expvalue})  above
increases,
 thereby enhancing the binding.

However, once $d \sim r_1$ the ``cages" trapping the particle within each
unit cell become tighter and the particle can only tunnel between the
different cells.
 The energy then rises and the bound state disappears.

 Finding the optimal $d$ (or density of the droplet $n$)
 and the corresponding  binding: $[E/N](n)$ at this number density
 requires detailed calculations beyond what we have attempted
here\cite{NandNinprep}.

%INSERT follows; Also Appendix I is new (8-17-04)

 Showing that one extra particle can be bound in a periodic  box of
size $L$ where the
 previous $N$ are located at prescribed positions $r_i^0$ (as we did
above
when the $r_i^0$
 were the nodes of a regular simple cubic crystal) is only {\it one} step
towards proving
 the existence of large Kaon droplets with the specific $ KK $
potential
above.  Indeed:

 i) While we optimize the inter-particle separation $d \sim
N^{(-1/3)}\cdot L $ to
minimize
$\epsilon_N$,  the energy of the $N+1^{th}$ particle added to the
lattice,  we
should verify
 that the same $d$ also allows each of the original $N$ particles to
bind
around
 the empty lattice site that it occupies.

 ii) We should show that the particular frozen $r_i^0$ arrangement at
{\it all} lattice
 sites within the periodic $L^3$ box represents the ``worst case" for
binding the $N+1^{th}$
 particle and any rearrangement of the $N$ bosons inside the box allows
stronger
 binding of the $N+1^{th}$ boson.

 iii) Using i) and ii) above we then show that enlarging the system
from $N$ to $ N+1$
 bosons {\it and} simultaneously letting the cube size adjust to the
new optimal
 length, $L(N) \rightarrow L(N+1)$, the energy is lowered by {\it more
} than the above
 binding:
\begin{equation}
 E[(N+1),L(N+1)] < E[N,L(N)] +  \epsilon_N
\label{newoptimallength}       %Eq (23) {*}
\end{equation}
This last step is readily achieved by employing
\begin{eqnarray}
E[(N+1),L(N+1)] = \nonumber
%& = & \langle \Psi(N+1)|H|\Psi(N+1) \rangle \nonumber \\
%& = & \int{\prod}_{i=1,2...N} d^3 r_i^0 \int{d^3 r_{(N+1)}}
%\Psi^*(r_{(N+1)};r_i^0) \nonumber \\
%& = &  
\\ \int \prod d^3 r_i^0 \int d^3 r_{N+1} \Psi(r_{N+1};r_i^0) \nonumber \\ 
\! \! \! \! \! \! \! \! \! \! \! \! \! \! \! \!\! \! \! \! \! \! \! \! \! \! \! \! \! \! \! \!\! \! \! \! \! \! \! \! \! \! \! \! \! \! \! \!\! \! \! \! \! \! \! \! \! \! \! \! \! \! \! \!\! \! \! \! \! \! \! \! \! \! \! \! \! \! \! \!\! \! \! \! \! \! \! \! \! \! \! \! \! \! \! \!\! \! \! \! \! \! \! \! \! \! \! \! \! \! \! \!\! \! \! \! \! \! \! \! \! \! \! \! \! \! \! \!\! \! \! \! \! \! \! \! \! \! \! \! \! \! \! \!\! \! \! \! \! \! \! \! \! \! \! \! \! \! \! \!\! \! \! \! \! \! \! \! \! \! \! \! \! \! \! \!\! \! \! \! \! \! \! \! \! \! \! \! \! \! \! \!\! \! \! \! \! \! \! \! \! \! \! \! \! \! \! \!\! \! \! \! \! \! \! \! \! \! \! \! \! \! \! \!\! \! \! \! \! \! \! \! \! \! \! \! \! \! \! \!\! \! \! \! \! \! \! \! \! \! \! \! \! \! \! \!
\times [ (\sum_{i=1,...N} - \frac{\hbar^{2}}{2m} \nabla^{2}_i 
+ \sum_{i>j=1...N}
V(r_i^0,r_j^0)) \nonumber \\
\! \! \! \! \! \! \! \! \! \! \! \! \! \! \! \!\! \! \! \! \! \! \! \! \! \! \! \! \! \! \! \!\! \! \! \! \! \! \! \! \! \! \! \! \! \! \! \!\! \! \! \! \! \! \! \! \! \! \! \! \! \! \! \!\! \! \! \! \! \! \! \! \! \! \! \! \! \! \! \!\! \! \! \! \! \! \! \! \! \! \! \! \! \! \! \!\! \! \! \! \! \! \! \! \! \! \! \! \! \! \! \!\! \! \! \! \! \! \! \! \! \! \! \! \! \! \! \!\! \! \! \! \! \! \! \! \! \! \! \! \! \! \! \!\! \! \! \! \! \! \! \! \! \! \! \! \! \! \! \!\! \! \! \! \! \! \! \! \! \! \! \! \! \! \! \!\! \! \! \! \! \! \! \! \! \! \! \! \! \! \! \!\! \! \! \! \! \! \! \! \! \! \! \! \! \! \! \!\! \! \! \! \! \! \! \! \! \! \! \! \! \! \! \!\! \! \! \! \! \! \! \! \! \! \! \! \! \! \! \!\! \! \! \! \! \! \! \! \! \! \! \! \! \! \! \!- \frac{\hbar^{2}}{2m}
\nabla^{2}_{(N+1)} + \sum_{i=1...N}V(r_i^0,r_{(N+1)}) ] \nonumber
\\  \Psi{(r_{N+1};r_i^0)}.
\label{laststep}       %Eq (24)
\end{eqnarray}

 We use the $^0$ suffixes on the first $N$ coordinates to emphasize
that in
evaluating
 the expectation of the kinetic energy and $N$ interactions of the
$N+1^{th}$
particle by
 doing the innermost $r_{(N+1)}$ integration, these first $N $
coordinates are
``frozen".
 This integral then yields $\epsilon [r_1^0,..,r_N^0]$, the binding
energy
of the extra
 particle to the first frozen $N$ which is then further averaged over
all
$r_i^0$ using
 the normalized measure provided by the density function of the first
$N$
particles.
 If (ii) holds this yields a value smaller than than $\epsilon_N$ above
corresponding
 to  the special case of a perfect full lattice of size $ L^3(N+1)$.

 Next consider the first term in the last integral, namely,  $H(N)$,
the
part of the
 Hamiltonian pertaining to the first $N$ bosons. Since $H(N)$ does {\it
not} depend on
$r_{(N+1)}$, the Normalized integral over $H(N)$ simply yields
$E[N,L(N+1)]$
which exceeds
$ E[N,L(N)]$. Adding the two terms we find at the desired inequality
(\ref{newoptimallength}).  %{*}

 Summing these over $n<N$ we yield a lower bound on the binding of the
droplet;
\begin{equation}
 E_N < N {\epsilon}_N
\label{lowerboundyield}   %Eq (25){**}
\end{equation}

 For a potential which is an attractive constant apart from hard core
cubes of size
$ 2r_1$ around each boson, i.e., the case studied above this last
inequality,
is directly
 proven in one dimension in Appendix I.  Unfortunately, this particular
elegant method
 does not readily generalize to three dimensions.

 Coming back to issues i) and ii), clearly filling up completely a
simple cubic
 lattice with the previous $N$ bosons- rather than leaving one ``hole"
i.e., a vacant
 lattice site where the extra particle nicely fits lowers the
volume of the ``cage" in which the $N$th particle is free to roam. 
This, in turn, elevates its kinetic energy.
 The attractive potential energy  is $ \{8 V[(\sqrt{3}/2) d] + 24
V[(\sqrt{11}/2)d] +... \}$.

Compare this to the  ``optimal" regular arrangement with one vacancy
in the regular
 cubic lattice.  The free volume where no strong repulsion occurs is
now $\sim$ twice as
 large making roughly for $2^{-(2/3)} \sim .65$ times lower kinetic
energy,
whereas the
 attractive potential here is $6 V[d] +12V[d\sqrt{2} ] +... $.
For $d \sim r_2 \sim 2m \sigma^{-1}$ the
 latter is similar to the potential energy in the previous case.
Hence we find that
 certain $d$ values which  allow binding of the $N+1^{th}$ particle to
a
perfect lattice
 make for even stronger binding inside the lattice.

 Next let us consider a random rearrangement of the $r_i^0$ inside the
$L^3$ cube with the
 same average number density R$d^{-3}$. It will contain pairs, triplets
etc., of particles
 which are  nearer to each other than $d$ and this will be compensated
by having
 nearby ``lacunas" of under-dense points.  The latter constitute ideal
placements of
 the extra $N+1^{th}$ particle: It will have more free space to move
and at
the same time
 will be more strongly attracted to the dense ``clusters".  Note that
in
evaluating
 $\epsilon [ r_1^0,r_2^0,...,r_N^0]$ the binding of the extra
$N+1^{th}$
particle to the
 frozen $N$ particles at $r_i^0$, we need {\it not} worry about the
fact
that the
 ``crowding" of some of the frozen vertices raises {\it their} mutual
interaction
 energies .

%     V.
\subsection{Some Concluding Remarks}

  In this paper we have shown a close correspondence between the field
theoretic
  concept of Q balls and droplets of coherent NR bosonic matter and
elaborated
  at some length on the criteria, in a NRS picture, for forming the
such droplets.

  While recently field theoretic/effective Lagrangian
methods---such as those used by Coleman in predicting Q balls---are
broadly applied to
 many-body physics, we find that the traditional variational NRS
approaches can be used to prove the existence of some ``particle
physics" Q balls.
 Unfortunately these Q balls with Q = Strangeness, Charm and Beauty,
while stable against
 decays via strong interactions do decay rather quickly via weak
interactions.
 A single $K^0$ decays in $\sim 10^{-10}$ sec (and $D^0, B^0$ decay 100
times faster).
 A droplet of $N$ non-relativistic, weakly
 bound neutral bosons will start disintegrating after times which
 are $1/N$ shorter than the decay time of a single boson. If the
minimal
 number of $K^0$'s can be viewed as an $ N \rightarrow \infty$
``droplet" is $\sim $ 100,
 we need to assemble within $\sim$ (5 Fermi)$^3$  100 slow $K^0$'s in a
 picosecond!, which seems impractical.

 We would like to note, however, that the general features of the $KK$
potential
facilitating droplet formation, namely, attraction at ``long" $\sim$ 0
(Fermi)
distances and repulsion at ``short" 1/3 Fermi distances, hold also for
the
$K$-Nucleon (and the $N-N$!) system.  While there are both few body and
``Droplets" of
 nucleons (a.k.a. ``Stable Nuclei") no $K-N$  bound states exist and the
$K-N$ S-wave
 scattering length is known to be repulsive.

 However $K^0$ droplets exist despite the absence of a $KK$ bound
 state. Could a $K^0$ which is free to move in a pre-existing large
 nucleus (so long as it avoids getting too close to the
 nucleons) bind to the latter? This is clearly not evident since
 the nuclear density and size are fixed by  nucleon-nucleon
 interaction and {\it not} by $K-N$ physics.

 It is amusing to note, however, that the existence of such states
 would manifest experimentally as follows:
Let a beam of slow $K^+$ charge exchange on a heavy nucleus with
the resulting $K^0$ binding to the nucleus.  The subsequent $K^0
\rightarrow  \pi^+ \pi^-$ decay $\sim  10^{-10}$ sec later is
likely to break the nucleus, yielding a spectacular ``Star" which,
in the absence of a $K^0$-nucleus bound state, should not occur.

\section{Appendix I}

 In the ``frozen $N$" variant, the $N+1^{th}$
 particle is restricted to the $N$ intervals of size $L^\prime/N$
between pairs
of existing
 particles (periodic boundary conditions avoid half size end
intervals) with
$ L'=L-2Nr_1$ the effective length allowed by the constraints.
 There is no tunneling between these $N$ intervals and the energy is :
\begin{equation}
          \epsilon_N =\hbar^2/[2m (L'/N)^2]
\label{epsilonenergy}  %(eq 26)
\end{equation}
  Note that $N \epsilon_N$ is the energy of {\it one} fictitious
representative
 particle moving in an {\it N } dimensional cube of side $L'/N$.
 The no-tunneling rigidity which is an artifact of one dimension,
reflects
 also in the full $N$-body problem of finding the ground state of the
Hamiltonian
 $- \frac{\hbar^{2}}{2m} \sum \frac{\partial^2}{\partial x_i^{2}}$ 
for $N$ particles in the $(0,L')$ interval
via the impenetrability---or ordering condition---
\begin{equation}
           0 < x_1 < x_2 <....x_N < L'.
\label{Def(P_N)}  %  \{ Def (P_N) }  (eq. 27)
\end{equation}
 Equivalently we have one particle restricted to the above $N $
dimensional parallelepiped 
$(P_N)$ satisfying the free Schr\"odinger equation in $N$
dimensions.

 (The $N!$ parallelepipeds 
$P_N$'s obtained by permuting the $x_i$ have the same volume
$L'^N/N!$
 and their union constitutes a cube of side $L'$). $P_N$ contains 
the $N$-dimensional cube $C_N$ of side $L/N$ :
\begin{eqnarray}
 0<x_1<L/N , \nonumber
\\ L/N <x_2 < 2L/N,
\nonumber
\\ ...,
\nonumber
\\ (k-1)L/N < x_k < kL/N, \nonumber
\\ ...,
\nonumber
\\ (N-1)L/N<
x_N < L , \nonumber
\\ \mbox{~and~} ||P_N|| > ||C_N||.
\label{ContainmentRelation1dim} % Eq (28)
\end{eqnarray}
 The variational principle and the above containment relation imply a
lower energy
 for the full problem.

\subsection{Acknowledgments}
 One of us (S.N.) would like to thank Steve Wallace for
 emphasizing the nuclear physics analogy.


\begin{thebibliography}{99}
\bibitem{HolygrailFN} Deriving from a microscopic theory of nuclear
forces $[E/N](n)$ which
correctly predicts the density and binding per particle of nuclear
matter has been the
 ``Holy grail" of extensive related research (originating with Bethe
and Pandharipande).
\end{thebibliography}
\end{document}